\documentclass[preprintnumbers,twocolumn,amsmath,amssymb,nofootinbib]{revtex4}

\usepackage{lineno,hyperref,amsmath}
\usepackage{graphicx,subcaption,epstopdf}
\usepackage{color}

\begin{document}

\title{Study of fully coupled 3D envelope instability using automatic differentiation}

\author{Ji Qiang}
\address{Lawrence Berkeley National Laboratory,
          Berkeley, CA 94720}

	

%
\begin{abstract}
Auto-differentiation is a powerful tool for computing derivatives of simulation results with respect to given parameters. In this letter, we have applied this tool to investigate the instability of a dynamics system that is governed by $21$ ordinary differential equations. This second-order
instability (named envelope instability) is driven by space-charge effects and has significant impact on the operational regimes of particle accelerators.
Our study delves into the three-dimensional envelope instability, incorporating both transverse and longitudinal coupling. Conventionally, analyzing this complex system would necessitate solving $441$ ordinary differential equations, which is computationally intractable. However, by employing auto-differentiation, we were able to track only $21$ equations. This approach allowed us to uncover an additional instability stopband, which arises from space-charge-induced coupling and has not been reported in previous studies.
This research highlights the significant advantages of auto-differentiation in analyzing complicated dynamical systems involving a large number of ordinary differential equations.

\end{abstract}



\maketitle

Automatic differentiation (AD) is a mathematical technique that enables the efficient computation of derivatives of complex functions with respect to a given set of parameters without numerical approximation or symbolic differentiation. This method has been widely adopted in the artificial intelligence and machine learning (AI/ML) community for training neural network parameters~\cite{ad,pytorch,tensorflow}. More recently, AD has found applications in the particle accelerator physics community, where it has been used to evaluate the sensitivity of simulation results to accelerator lattice parameters and to accelerate optimization procedures~\cite{roussel2022,roussel2023a,roussel2023b,qiang2023,cheetah,wan}.
However, to the best of our knowledge, AD has not previously been applied to dynamical systems for stability analysis. Traditionally, stability studies of dynamical systems involve solving the transfer matrix equations for perturbed tangent vectors of the system variables. For highly complex systems, such as three-dimensional (3D) envelope evolutions with coupling induced by space-charge effects, the number of ordinary differential equations (ODEs) in the tangent vector transfer matrix can become prohibitively large—often exceeding 400—making direct computation intractable.
In such cases, AD offers an efficient alternative. By solving only the original system of ODEs—typically comprising about 20 equations—AD can automatically generate the corresponding transfer matrix for the tangent vectors. This significantly reduces computational complexity while preserving accuracy, making AD a powerful tool for stability analysis in complex dynamical systems.

In the particle accelerator community, envelope instability in a periodic transport channel refers to the instability in the evolution of the second moments of a charged particle distribution as it propagates through the channel. This second-order collective instability, driven by direct space-charge effects, can lead to beam size blow-up, emittance growth, and potential particle losses, thereby significantly constraining the operational regime of high-intensity accelerators. This instability has been recognized as a critical limitation in intense beam transport and has been the subject of extensive investigation since the 1980s~\cite{ingo1,jurgen,ingo3,chernin,barnard,okamoto,fedotov0,fedotov,lund,
tiefenback,gilson,groening,qin,li0,li,ingo2,goswami,jeon2,oliver,ito,yuan,ingo4,ingoprab17,qiangenv,qiangenv2}.
Most of those studies have focused on two-dimensional coasting charged particle beams. More recently, an investigation of bunched beam instability in a periodic quadrupole RF transport channel, based on an upright 3D envelope model without space-charge induced coupling from {initial} rotation, revealed a significant influence of longitudinal synchrotron motion on instability stopbands~\cite{qiangenv}.  
In this study, we examine a 3D bunched beam propagating through a periodic focusing lattice in an accelerator operating without acceleration. 
This scenario, where no external skew or tilt focusing is present and the only coupling arises from space-charge forces associated with beam rotation, has not been studied before.
Following the notation of Ref.~\cite{ryne}, for a charged particle in a periodic focusing lattice without acceleration, the Hamiltonian in terms of the six dimensionless phase-space coordinates  
$\zeta = (\bar{x}, \bar{p}_x, \bar{y}, \bar{p}_y, \bar{z}, \bar{p}_z)$
can be expressed as:
\begin{eqnarray}
H(\zeta,s)  & = &
\frac{\delta}{lp_0}(\frac{\bar{p}_x}{2} +\frac{\bar{p}_y}{2}+\frac{\bar{p}_z}{2})+\frac{qg(s)l}{\delta}(\frac{\bar{x}^2}{2}-\frac{\bar{y}^2}{2})+ \nonumber \\
& &
   \frac{l}{\delta} k^2_z(s) \frac{{\bar z}^2}{2}+\frac{p_0}{\delta l}K \phi(\bar{x},\bar{y},\bar{z})
\end{eqnarray}
where $\bar{x} = x / l$, $\bar{y} = y / l$, and $\bar{z} = \Delta z\,\gamma / l$ are the normalized transverse positions and the relative longitudinal position with respect to the reference particle;  
$\bar{p}_x = p_x / \delta$, $\bar{p}_y = p_y / \delta$, and $\bar{p}_z = \Delta E / (p_0 c)$ are the normalized transverse momenta and the normalized longitudinal energy deviation with respect to the reference particle.  
Here, $\delta = m c$, where $m$ is the particle rest mass and $c$ is the speed of light in vacuum; $l = c / \omega$ is the scaling length, with $\omega$ the reference angular frequency; $\gamma = 1 / \sqrt{1 - \beta^2}$ is the relativistic Lorentz factor, with $\beta = v / c$ and $v$ the speed of the reference particle; $p_0$ is the reference particle momentum; $q$ is the particle charge; $g$ is the quadrupole gradient; $k_z$ is the longitudinal focusing strength;  
$K = \frac{q}{m c^2 \gamma^2 \beta^2}$ is the space-charge factor; and $\phi$ is the space-charge potential from a uniform ellipsoidal density distribution, given by  
\begin{equation}
\phi = \frac{Q}{4\pi \epsilon_0} \frac{\lambda_3}{l} 
\left[ -\frac{1}{2} 
(\bar{x}^2, \bar{y}^2, \bar{z}^2) \,
A G A^{T} \,
(\bar{x}^2, \bar{y}^2, \bar{z}^2)^{T} 
\right],
\end{equation}
where $G$ is a $3\times 3$ diagonal matrix with nonzero diagonal elements $(g_{311}, g_{131}, g_{113})$, obtained from the eigenvalues of the $3\times 3$ covariance matrix of the spatial coordinates;  
$A$ is the eigenvector matrix associated with these eigenvalues; and the superscript $T$ denotes matrix transpose.  
The constant $\lambda_3$ depends weakly on the beam distribution: $\lambda_3 = \frac{1}{5\sqrt{5}}$ for a uniform distribution, $\lambda_3 = \frac{1.01}{5\sqrt{5}}$ for a parabolic distribution, and $\lambda_3 = \frac{1.05}{5\sqrt{5}}$ for a Gaussian distribution~\cite{sacherer}.
The diagonal elements $g_{lmn}$ are given by  
\begin{equation}
    g_{lmn}(x, y, z) = \frac{3}{2} \int_{0}^{\infty} 
    \frac{ds}{(x^{2} + s)^{l/2} (y^{2} + s)^{m/2} (z^{2} + s)^{n/2}},
\end{equation}
where $x$, $y$, and $z$ are the square roots of the three eigenvalues of the spatial position covariance matrix.

Given the above Hamiltonian, the equations of motion for the particle coordinates can be written as  
\begin{equation}
    \frac{d\zeta}{ds} = F(s)\,\zeta,
\end{equation}
where $F(s)$ is a $6 \times 6$ matrix given by
\begin{eqnarray}
	F(s) & = & \begin{pmatrix}
		0 & \frac{\delta}{lp_0} & 0 & 0 & 0 & 0 \\
		F_{21}(s) & 0 & F_{23}(s) & 0 & F_{25}(s) & 0 \\
		0 & 0 & 0 &  \frac{\delta}{lp_0} & 0 & 0 \\
		F_{41}(s) & 0 & F_{43}(s) & 0 & F_{45}(s) & 0 \\
		0 & 0 & 0 & 0 & 0 & \frac{\delta}{lp_0} \\
		F_{61}(s) & 0 & F_{63}(s) & 0 & F_{65}(s) & 0 \\
		 \end{pmatrix}
\end{eqnarray}
The elements $F_{ij}(s)$ are given by  
{\small
\begin{eqnarray}
	F_{21}(s) & = & -\frac{q g(s)l}{\delta} + \bar{K} (g_{311}A_{11}^2+
    g_{131}A_{12}^2+g_{113}A_{13}^2)
	 \\
F_{23}(s) & = & \bar{K} (g_{311}A_{11}A_{21}+
    g_{131}A_{12}A_{32}+g_{113}A_{13}A_{33}) \\
F_{25}(s) & = & \bar{K} (g_{311}A_{11}A_{31}+
    g_{131}A_{12}A_{22}+g_{113}A_{13}A_{23}) \\
    F_{41}(s) & = & F_{23}(s)  \\
	F_{43}(s) & = &  \frac{q g(s)l}{\delta} + 
  \bar{K} (g_{311}A_{21}^2+
    g_{131}A_{22}^2+g_{113}A_{23}^2)  \\
F_{45}(s) & = &\bar{K} (g_{311}A_{21}A_{31}+
    g_{131}A_{22}A_{32}+g_{113}A_{23}A_{33}) \\
F_{61}(s) & = & F_{25}(s)  \\
F_{63}(s) & = & F_{45}(s) \\
	F_{65}(s) & = &  -\frac{k_z^2l}{\delta} + 
    \bar{K} (g_{311}A_{31}^2+
    g_{131}A_{32}^2+g_{113}A_{33}^2)
\end{eqnarray}
}
and $\bar{K}=\frac{Q}{4\pi \epsilon_0}\frac{\lambda_3}{l}\frac{p_0}{\delta l}K$.
From the particle coordinate evolution equation, the evolution of the covariance matrix of the particle distribution,  
$\Sigma_{ij} = \langle \zeta_i \zeta_j \rangle - \langle \zeta_i \rangle \langle \zeta_j \rangle$,
can be expressed as  
\begin{equation}
    \frac{d\Sigma}{ds} = F\,\Sigma + (F\,\Sigma)^{T},
    \label{covarEq}
\end{equation}
where $\langle \cdot \rangle$ denotes an average over the particle distribution in six-dimensional phase space.  
The covariance matrix $\Sigma$ is symmetric and semi-positive definite, with $21$ independent components.  
Therefore, Eq.~\eqref{covarEq} represents a coupled system of $21$ independent ordinary differential equations.

To study the stability of the above equations for a periodic system, the conventional approach is to introduce a perturbation to the original envelope equation,  
\[
\Sigma(s) = \Sigma_m(s) + \tilde{\Sigma}(s),
\]
where $\Sigma_m(s)$ is the matched solution to the envelope equation, and $\tilde{\Sigma}(s)$ (also called the \emph{tangent vector}) denotes the perturbation to the matched solution.  
The evolution of the tangent vector is governed by another set of $21$ ordinary differential equations.  
To analyze the stability of the evolution, the $21$ independent perturbation components can be assembled into a vector  
$\xi = \left( \tilde{\Sigma}_{11}, \tilde{\Sigma}_{12}, \ldots, \tilde{\Sigma}_{66} \right)^{T}$.
The equations of motion for the perturbations are then given by  
\begin{equation}
    \frac{d\xi}{ds} = D(\Sigma_m,s)\,\xi(s),
    \label{varEq}
\end{equation}
where $D(s)$ is the $21 \times 21$ perturbation matrix.  

Let the solution to the above equation be written as $\xi(s) = M(s)\,\xi(0)$, where $M(s)$ is the $21 \times 21$ transfer matrix of the perturbations.  
Substituting into Eq.~\eqref{varEq} yields  
\begin{equation}
    \frac{dM(s)}{ds} = D(\Sigma_m,s)\,M(s),
    \label{tranEq}
\end{equation}
with $M(0)$ equal to the $21 \times 21$ identity matrix.  
This system consists of $441$ coupled ordinary differential equations, which can be solved numerically using the matched envelope solution $\Sigma_m(s)$.  

The stability of the envelope perturbations is determined from the eigenvalues of the transfer matrix $M(L)$ over one lattice period $L$.  
For the envelope oscillations to remain stable, all $21$ eigenvalues of $M(L)$ must lie on the unit circle in the complex plane.  
The magnitude of each eigenvalue gives the envelope mode growth (or damping) rate per lattice period, while its phase corresponds to the mode oscillation frequency.  
If the magnitude of any eigenvalue exceeds unity, the corresponding envelope mode becomes unstable.

The direct numerical solution of the $441$ coupled ODEs in Eq.~\eqref{tranEq} can be complicated and computationally intractable.  
Instead, we utilize the solution of the covariance matrix equation, Eq.~\eqref{covarEq}, after one lattice period, which can be expressed as a function of the initial condition:
\begin{equation}
    \Sigma(L) = f\!\left( \Sigma_0 \right).
\end{equation}
To first order in the deviation from the matched solution, this can be written as
\begin{equation}
    \Sigma(L) \approx f\!\left( \Sigma_{m0} \right) 
    + \frac{\partial f}{\partial \Sigma_{m0}} 
    \left[ \Sigma_0 - \Sigma_{m0} \right],
\end{equation}
which leads to
\begin{equation}
    M(L) = \frac{\partial \Sigma_L}{\partial \Sigma_{m0}}.
\end{equation}

The above equation shows that the Jacobian matrix, formed from the first-order derivatives of the covariance matrix elements after one lattice period with respect to their initial values, yields the transfer matrix of the perturbation vector.  
These derivatives can be computed to machine precision using the automatic differentiation (AD) technique.

In this study, we adopt a forward-mode method based on the first-order truncated power series algebra (TPSA)~\cite{qiangipac25}, also known as the dual-number representation~\cite{adwiki}.  
This method transforms the computation of a function’s derivatives with respect to its variables into the evaluation of the function using a vector-like differentiable variable, according to specific algebraic rules.  
Here, the differentiable vector variable $F$ is defined as  
$F = \left( f, f_{x_1}, f_{x_2}, \ldots, f_{x_n} \right)$,
where the first element is the function value, and the subsequent elements are the partial derivatives with respect to each variable of interest.

The algebraic rules for two differentiable vector variables $F$ and $G$ are as follows:

Addition rule:
{\small
\begin{equation}
F + G = \left( f + g,\; f_{x_1} + g_{x_1},\; f_{x_2} + g_{x_2},\; \ldots,\; f_{x_n} + g_{x_n} \right)
\end{equation}
}
Multiplication rule:
{\small
\begin{equation}
F\,G = \left( f g,\; g f_{x_1} + f g_{x_1},\; g f_{x_2} + f g_{x_2},\; \ldots,\; g f_{x_n} + f g_{x_n} \right)
\end{equation}
}
Division rule:
{\small
\begin{equation}
\frac{F}{G} = \left( \frac{f}{g},\; \frac{f_{x_1} g - g_{x_1} f}{g^2},\; \frac{f_{x_2} g - g_{x_2} f}{g^2},\; \ldots,\; \frac{f_{x_n} g - g_{x_n} f}{g^2} \right)
\end{equation}
}
For a scalar function $h(F)$ of a differentiable vector variable $F$, we have:
\begin{equation}
h(F) = \left( h(f),\; h_f\,f_{x_1},\; h_f\,f_{x_2},\; \ldots,\; h_f\,f_{x_n} \right),
\end{equation}
where $h_f$ denotes the partial derivative of $h$ with respect to $f$, i.e., $h_f = \partial h / \partial f$.

Based on the computational rules described above, we define a new data type for the differentiable vector in the { \texttt{Fortran90}} programming language, together with its corresponding operators and common special functions (e.g., exponential and trigonometric functions).  
These differentiable variables are then used to solve Eq.~\eqref{covarEq} numerically in the same manner as conventional double-precision variables, employing a fourth-order Runge-Kutta integration scheme. {
In contrast to a conventional computer program where the $\Sigma$ variables are declared as double precision, the auto-differentiable program declares them as differentiable variables. While the data type is different, the mathematical expressions and the numerical algorithm remain identical. The initial values of the $\Sigma$ variables are defined as a set of independent differentiable variables $(x_1, x_2, \cdots x_{21})$. As the integration proceeds, these $\Sigma$ variables evolve as functions of their initial conditions. Consequently, upon completion of the integration, the final $\Sigma$ variables contain not only their function values but also their derivatives with respect to those initial values.}
The transfer matrix is then directly constructed from these derivatives.
\begin{figure}[!htb]
\includegraphics[width=8cm]{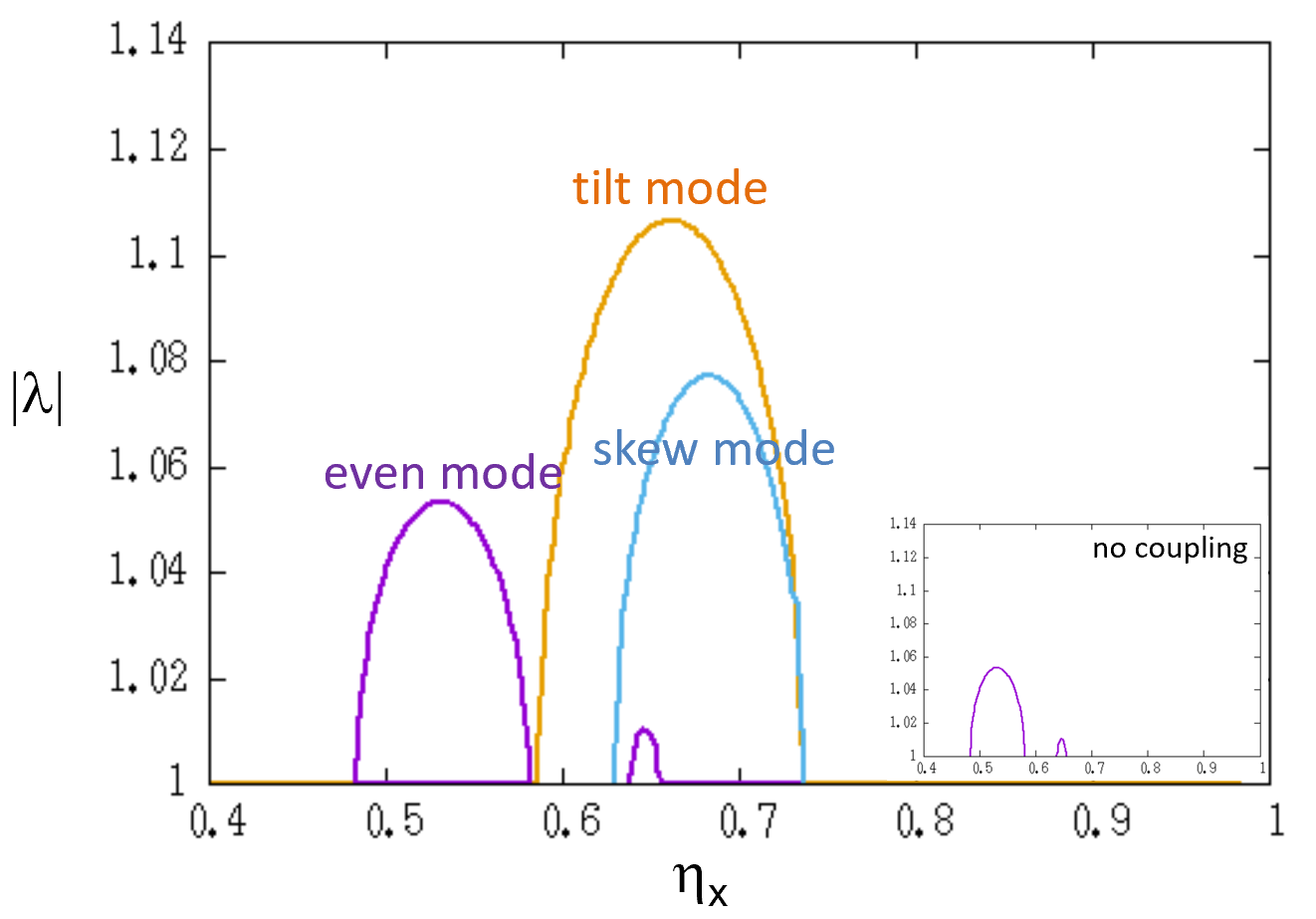}
\caption{
Growth rate amplitudes of unstable envelope modes as a function of the {horizontal tune depression}, shown both without (embedded plot) and with six-dimensional (6D) coupling, for a longitudinal zero-current phase advance of $60^\circ$.
}
\label{fig3}
\end{figure}  

\begin{figure}[!htb]
\includegraphics[width=8cm]{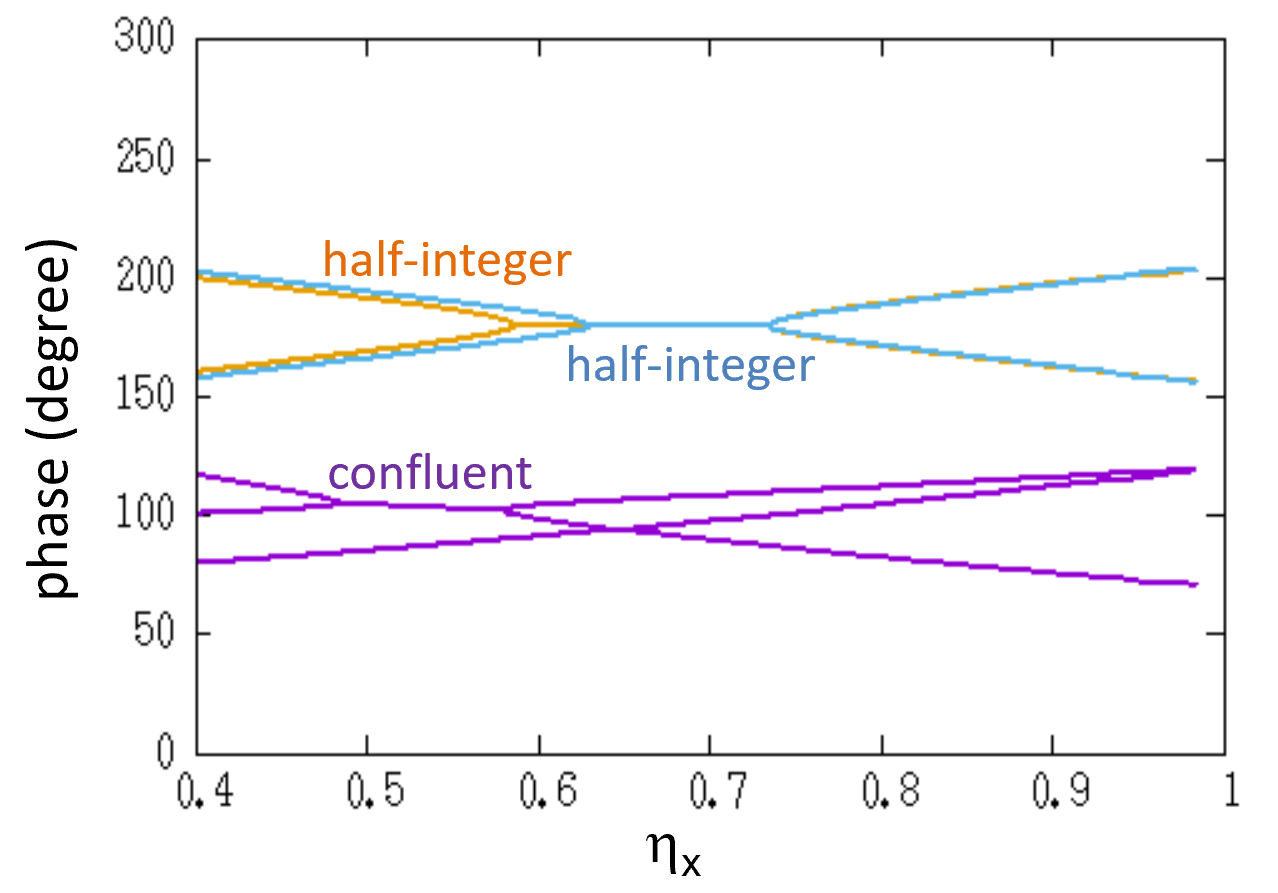}
\caption{
Phases of unstable envelope modes as a function of the { horizontal tune depression}, with coupling and a longitudinal phase advance of $60^\circ$.}
\label{fig4}
\end{figure}  
As an illustration, we consider a periodic lattice with transverse focusing provided by quadrupole magnets and longitudinal focusing from RF bunchers.  
For a proton beam with a kinetic energy of $150\,\mathrm{MeV}$, the zero-current phase advance per period is approximately $60^\circ$ in the horizontal ($x$) dimension and $145^\circ$ in the vertical ($y$) dimension.
Fig.~\ref{fig3} shows the envelope mode growth rate amplitude as a function of the { horizontal tune depression}, for a longitudinal zero-current phase advance of $60^\circ$.  
Without coupling, two instability stopbands are observed: one spanning a depressed phase advance from approximately $0.48$ to $0.58$, and another centered around $0.65$.  
When space-charge-induced coupling is included, two additional instability stopbands appear.  
One corresponds to the skew mode arising from coupling in the transverse ($x$--$y$) plane{~\cite{oliver,yuan}}, and the other corresponds to the tilt mode resulting from coupling between the transverse and longitudinal planes.  
These additional stopbands extend over a depressed phase advance range from $0.58$ to approximately $0.74$, and exhibit larger growth rate amplitudes than those observed in the uncoupled case.

To investigate the mechanism responsible for the instability stopbands, we also plot the unstable mode phases as a function of the {horizontal tune depression} in Fig.~\ref{fig4}.  
Without coupling, the instability arises when two modes have the same phase and couple with each other.  
This type of instability is known as \emph{confluent instability} or \emph{coupled-mode instability}.  
When space-charge-induced coupling is present, an additional mechanism emerges: the unstable skew/tilt mode phase locks at $180^\circ$.  
In this case, the envelope modes resonate with the lattice structure, leading to instability.  
This phenomenon is referred to as a \emph{half-integer parametric resonance}.

\begin{figure}[!htb]
\includegraphics[width=7cm]{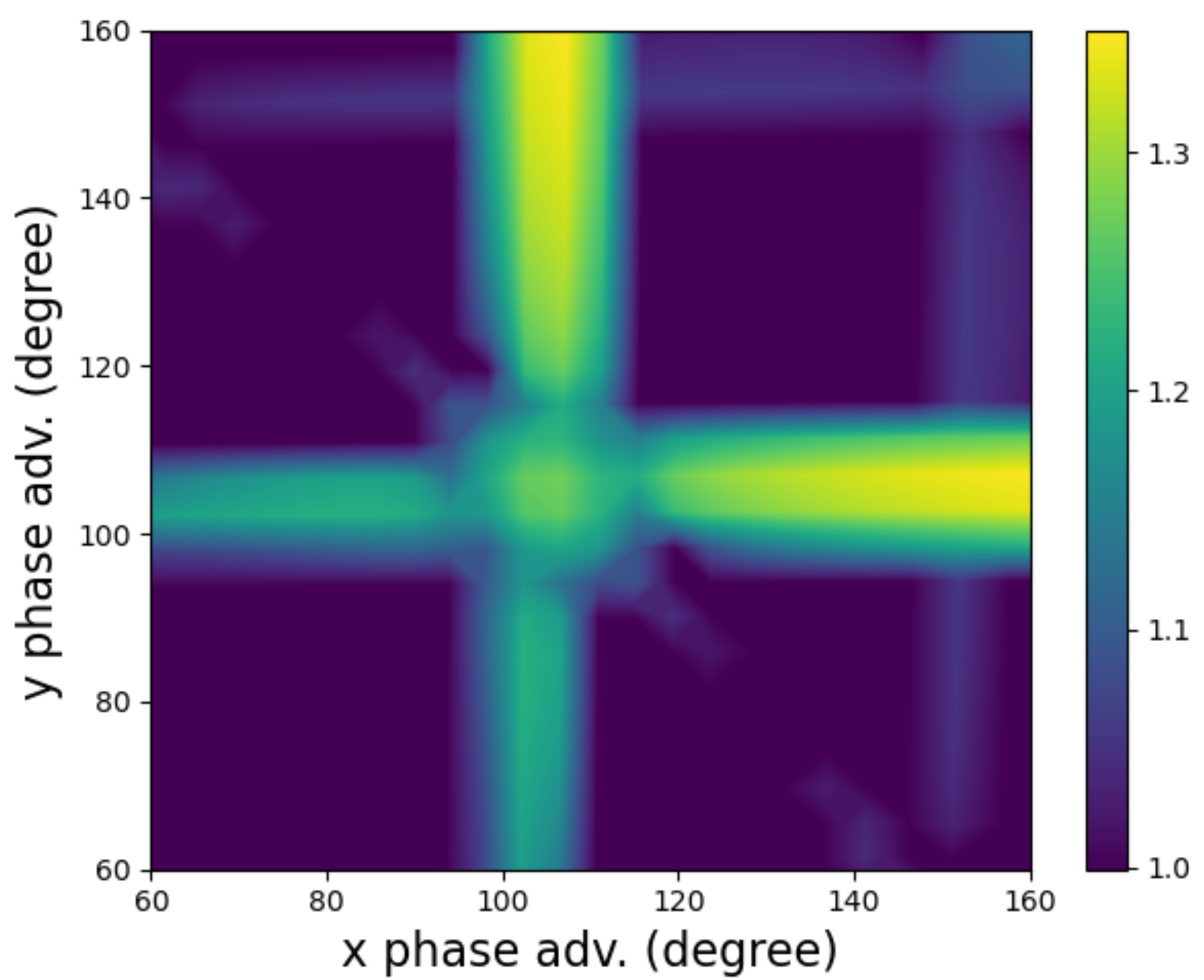}
\includegraphics[width=7cm]{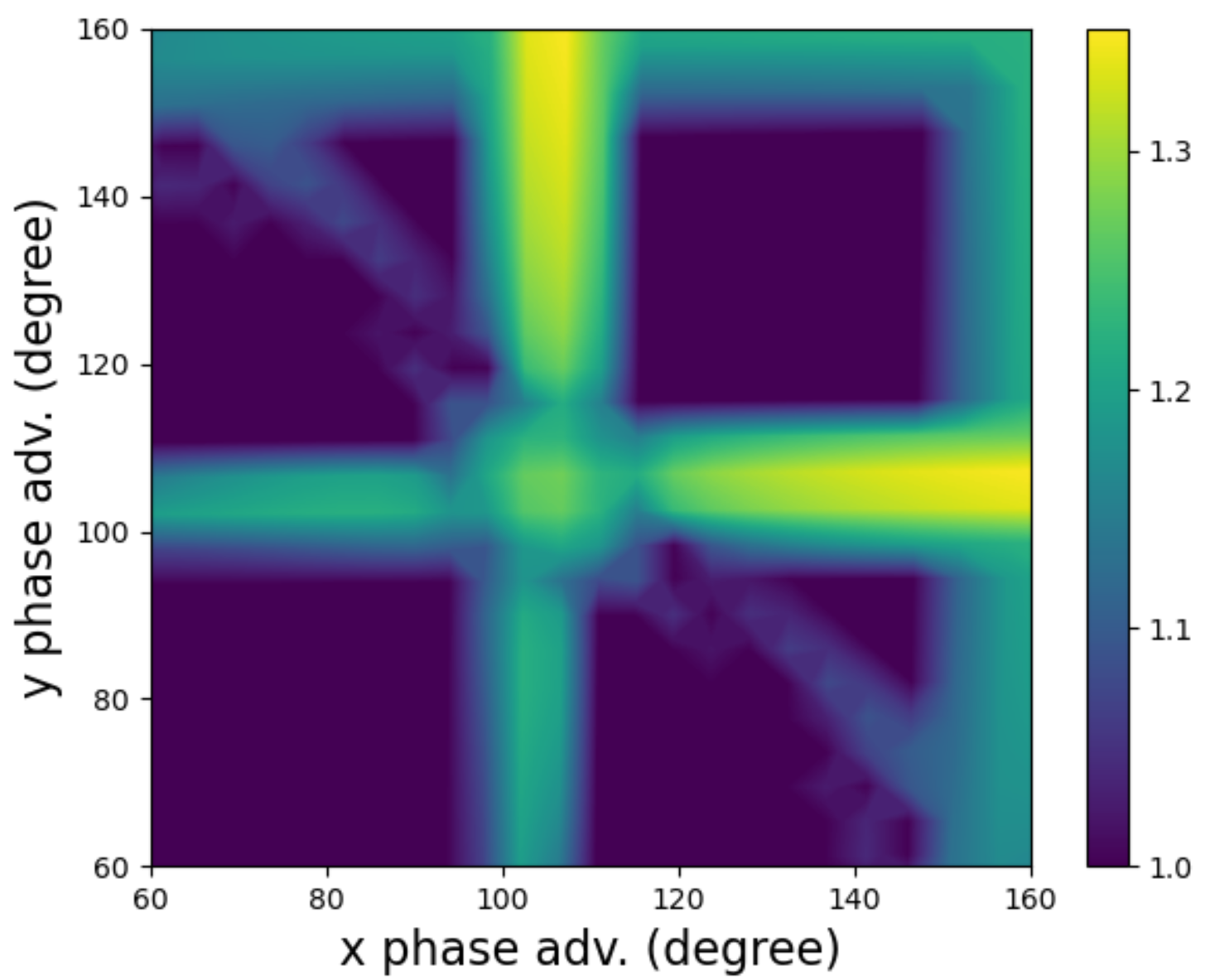}
\caption{
Scan of the transverse $x$- and $y$-zero-current phase advances showing the maximum envelope mode growth rate amplitudes without coupling (top) and with coupling (bottom).}
\label{fig5}
\end{figure}
To further explore the effects of coupling among the three dimensions, we performed two-dimensional zero-current phase advance scans in both the transverse ($x$--$y$) plane and the transverse-to-longitudinal plane, with and without coupling.  
Figure~\ref{fig5} shows the results for the maximum envelope mode growth rate amplitude from the $x$--$y$ plane scan, with the longitudinal zero-current phase advance fixed at $60^\circ$, a proton beam current of $40\,\mathrm{mA}$, and an RF bunching frequency of $650\,\mathrm{MHz}$.  
Two strong instability stopbands are observed near a $100^\circ$ zero-current phase advance in both the coupled and uncoupled cases, corresponding to half-integer resonance.  
However, when rotation-induced coupling is included, larger instability stopbands appear in the range of approximately $140^\circ$ to $160^\circ$ zero-current phase advance, as well as along the diagonal line in the scan.  
These coupling-induced stopbands are dominated by the half-integer resonance of the skew/tilt envelope modes.

\begin{figure}[!htb]
\includegraphics[width=6.9cm]{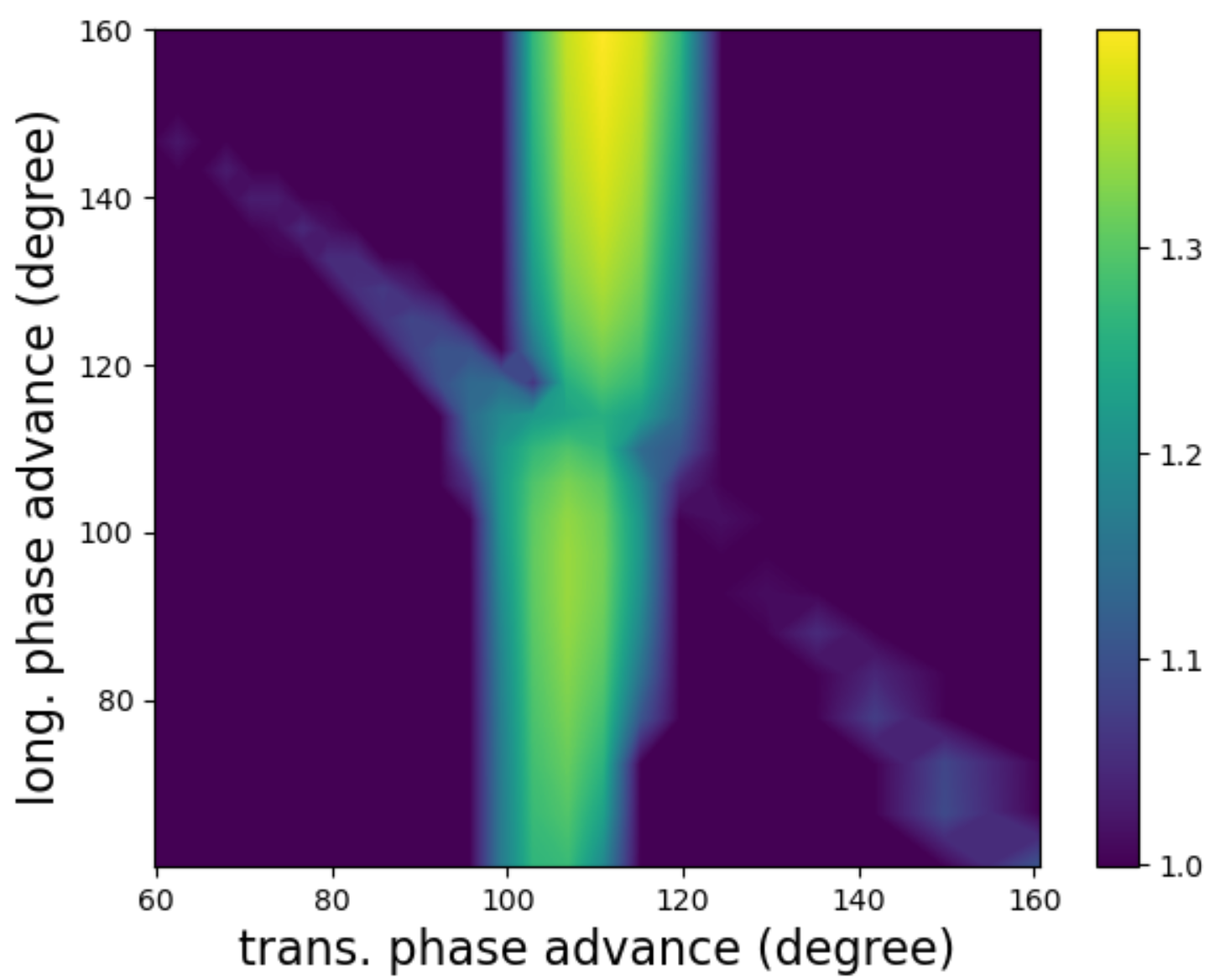}
\includegraphics[width=6.7cm]{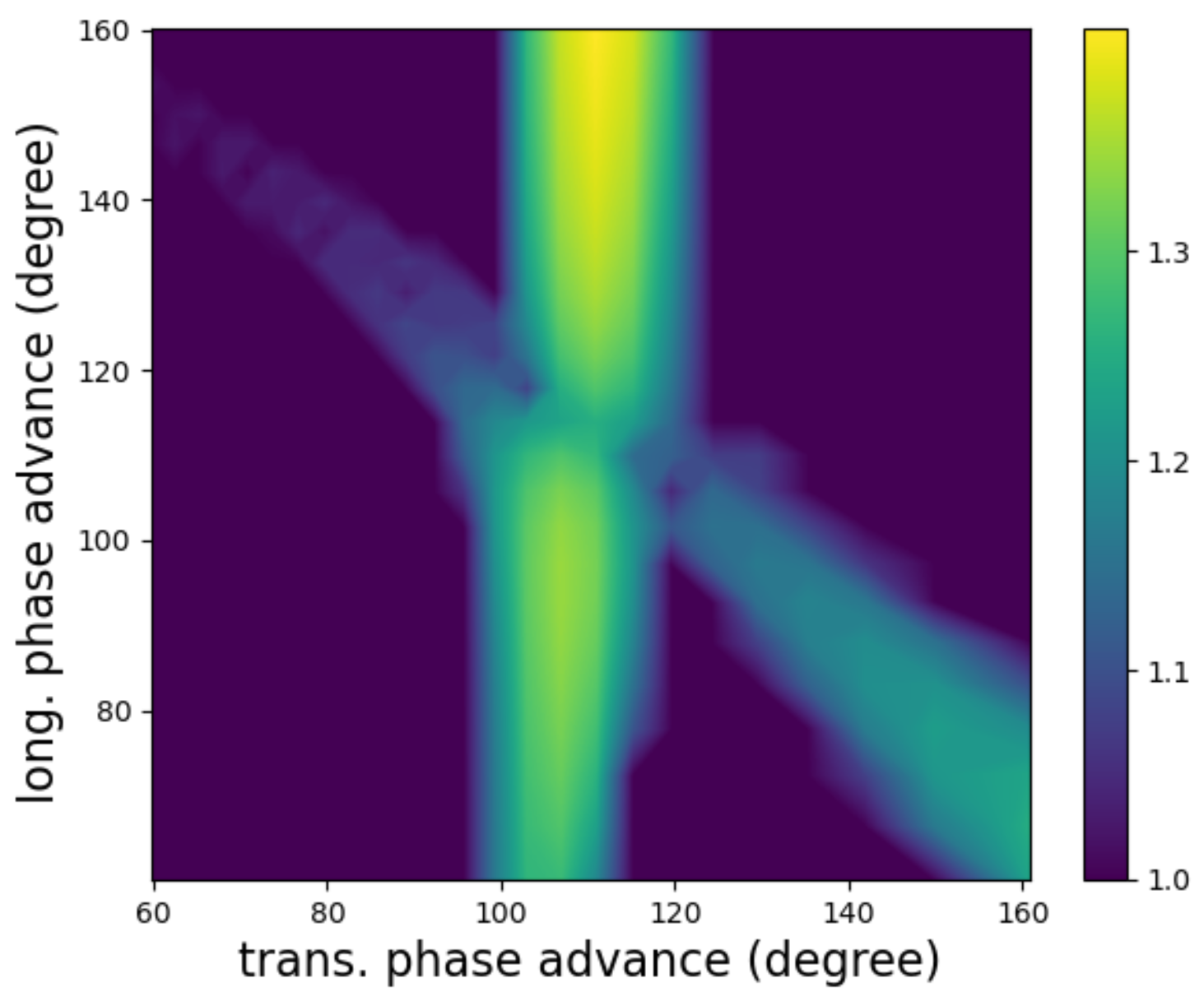}
\caption{
Scan of the transverse-to-longitudinal zero-current phase advances showing the maximum envelope mode growth rate amplitudes without coupling (top) and with coupling (bottom).}
\label{fig6}
\end{figure}
We also performed a two-dimensional zero-current phase advance scan between the transverse and longitudinal dimensions.  
In this case, the zero-current phase advances in the horizontal and vertical dimensions were kept equal.  
Figure~\ref{fig6} shows the maximum envelope mode growth rate amplitude as a function of the transverse and longitudinal zero-current phase advances.  
Without coupling, a major instability stopband appears around a $100^\circ$ transverse phase advance, corresponding to a half-integer resonance.  
This stopband is absent in the longitudinal dimension due to the presence of two identical RF focusing bunchers in a single lattice period.  
For a zero-current phase advance of about $100^\circ$ through the lattice period, the zero-current phase advance through a single RF buncher is less than $90^\circ$, preventing the occurrence of a half-integer resonance in the longitudinal plane.  
In the uncoupled case, a weak instability is observed along the diagonal line of the zero-current phase advance plane.  
With coupling, this instability region along the diagonal becomes broader and stronger, as the skew/tilt envelope modes driven by direct space-charge effects of rotated beam become unstable in this region.

In summary, we have shown that automatic differentiation (AD) can be effectively used to study the stability of complex dynamical systems, such as the dynamics of three-dimensional (3D) envelope modes of a charged particle beam.  
Application of this technique to the study of 3D envelope instability with full coupling has revealed new unstable regimes in the accelerator lattice parameter space that cannot be predicted by the original uncoupled upright 3D model.  


I would like to thank
Dr. R. D. Ryne for the use of his upright 3D envelope code. This research was supported by the
U.S. Department of Energy under Contract No. DE-AC02-05CH11231 and used computer
resources at the National Energy Research Scientific Computing (NERSC) Center.


	{%

} 
%
%

  

\begin{thebibliography}{9} 

\bibitem{ad}
C. C. Margossian, A Review of Automatic Differentiation and its Efficient Implementation,
Wiley interdisciplinary reviews: data mining and knowledge discovery 9, no. 4 (2019): e1305. 
\url{https://doi.org/10.1002/widm.1305}
\bibitem{pytorch}
A. Paszke et al., 
PyTorch: An Imperative Style,
High-Performance Deep Learning Library, in Advances
in Neural Information Processing Systems 32 , edited by
H. Wallach, H. Larochelle, A. Beygelzimer, F. d. Alche-Buc, 
E. Fox, and R. Garnett (Curran Associates, Inc.,2019) pp. 8024-8035.
\bibitem{tensorflow}
Martin Abadi et al., 
TensorFlow: Large-scale machine learning on heterogeneous systems,
2015.
\url{https://www.tensorflow.org/}   

\bibitem{roussel2022}
R. Roussel, A. Edelen, D. Ratner, K. Dubey, J. P. Gonzalez-Aguilera, Y.K. Kim, and N. Kuklev, 
Differentiable Preisach modeling for characterization and optimization of particle accelerator systems with hysteresis,
Phys. Rev. Lett. 128, 204801 (2022).
\bibitem{roussel2023a}R. Roussel and A. L. Edelen, 
Applications of differentiable physics simulations in particle accelerator modeling, arXiv:2211.09077.
\bibitem{roussel2023b}R. Roussel, A. Edelen, C. Mayes, D. Ratner, J. P. Gonzalez-Aguilera, S. Kim, E. Wisniewski, and J. Power, 
Phase space reconstruction from accelerator beam measurements using neural networks and differentiable simulations,
Phys. Rev. Lett. 130, 145001 (2023).
\bibitem{qiang2023}J. Qiang, 
Differentiable self-consistent space-charge simulation for accelerator design,
Phys. Rev. Accel Beams, vol. 26, 024601, 2023. \url{https://doi.org/10.1103/PhysRevAccelBeams.26.024601}
\bibitem{cheetah}J. Kaiser, C. Xu, A. Eichler, and A. S. Garcia,
Bridging the gap between machine learning and particle accelerator
physics with high-speed, differentiable simulations,
Phys. Rev. Accel Beams, vol. 27, 054601, 2024. 
\url{https://doi.org/10.1103/PhysRevAccelBeams.27.054601}
\bibitem{wan}J. Wan, H. Alamprese, C. Ratcliff, J. Qiang, Y. Hao, JuTrack: a Julia package for auto-differentiable accelerator modeling and particle tracking, 
Comp. Phys. Comm. 309, 109497 (2025).
 
\bibitem{ingo1}I. Hofmann, L. J. Laslett, L. Smith, and I. Haber, 
Stability
of the Kapchinskij-Vladimirskij (KV) distribution in long
periodic transport systems,
Part. Accel. 13, 145 (1983).
\bibitem{jurgen}J. Struckmeier and M. Reiser, Theoretical studies of
envelope oscillations and instabilities of mismatched intense
charged particle beams in periodic focusing channels,
Part. Accel. 14, 227 (1984).
\bibitem{ingo3}I. Hofmann, Stability of anisotropic beams with space
charge, Phys. Rev. E 57, 4713 (1998). 
\bibitem{chernin}D. Chernin, Evolution of rms beam envelopes in transport
systems with linear xy coupling, Part. Accel. 24, 29 (1988).
\bibitem{barnard}J. J. Barnard and B. Losic, Envelope modes of beams with angular momentum, in Proceedings of the 20th
International Linac Conference, Monterey, CA, 2000,
p. 293.
\bibitem{okamoto}H. Okamoto and K. Yokoya, Parametric resonances in
intense one-dimensional beams propagating through a
periodic focusing channel, Nucl. Instrum. Methods Phys. Res., A 482, 51 (2002). 
\bibitem{fedotov0}A. V. Fedotov and I. Hofmann, Half-integer resonance
crossing in high-intensity rings, Phys. Rev. Spec. Top.-Accel. Beams 5, 024202 (2002). 
\bibitem{fedotov}A. V. Fedotov, I. Hofmann, R. L. Gluckstern, and H. Okamoto,  Parametric collective resonances and spacecharge limit in high-intensity rings, Phys. Rev. Spec. Top.-Accel. Beams 6, 094201 (2003). 
\bibitem{lund}S. M. Lund and B. Bukh, Stability properties of the
transverse envelope equations describing intense ion beam
transport, Phys. Rev. Spec. Top.-Accel. Beams 7, 024801 (2004). 
\bibitem{tiefenback}M. Tiefenback, Space charge limits on the transport of ion beams in a long alternating gradient system, Ph.D. thesis, Lawrence Berkeley National Laboratory Report LBL-22465, 1986.
\bibitem{gilson}E. P. Gilson, M. Chung, R. C. Davidson  P. C. Efthimion, and R. Majeski, Transverse beam compression on the Paul
trap simulator experiment, Phys. Rev. Spec. Top.-Accel. Beams 10, 124201 (2007). 
\bibitem{groening}L. Groening, W. Barth, W. Bayer, G. Clemente, L. Dahl, P. Forck, P. Gerhard, I. Hofmann, M. S. Kaiser, M. Maier, S. Mickat, and T. Milosic,  Experimental evidence of the
90° Stop Band in the GSI UNILAC, Phys. Rev. Lett. 102, 234801 (2009). 
\bibitem{qin}H. Qin, M. Chung, and R. C. Davidson, 
Generalized Kapchinskij-Vladimirskij distribution and envelope equation
for high-intensity beams in a coupled transverse focusing lattice, Phys. Rev. Lett. 103, 224802 (2009).
\bibitem{li0}C. Li and Y. L. Zhao, Envelope instability and the fourth order
resonance, Phys. Rev. ST Accel. Beams 17, 124202 (2014).
\bibitem{li}C. Li and Q. Qin, Space charge induced beam instability in
periodic focusing channel, Phys. Plasmas 22, 023108 (2015).
\bibitem{ingo2}I. Hofmann and O. Boine-Frankenheim,  Space-Charge
Structural Instabilities and Resonances in High-Intensity
Beams, Phys. Rev. Lett. 115, 204802 (2015). 
\bibitem{goswami}A. Goswami, P. Sing Babu, and V. S. Pandit,
Beam
dynamics and stability analysis of an intense beam
in a continuously twisted quadrupole focusing channel, Eur. Phys. J. Plus (2016) 131: 393
\bibitem{jeon2}D. Jeon, J. H. Jang, and H. Jin, Interplay of space-charge
fourth order resonance and envelope instability, 
	Nucl. Instrum. Methods Phys. Res., A 832, 43 (2016). 
\bibitem{oliver}O. Boine-Frankenheim, I. Hofmann, and J. Struckmeier,
Parametric sum envelope instability of periodically focused intense beams,
Phys. Plasmas 23, 090705 (2016).

\bibitem{ito}K. Ito, H. Okamoto, Y. Tokashiki, and K. Fukushima, 
Coherent resonance stop bands in alternating gradient beam
transport, Phys. Rev. Accel. Beams 20, 064201 (2017).
\bibitem{yuan}Y. Yuan, O. Boine-Frankenheim, and I. Hofmann,
Modeling of second order space charge driven coherent
sum and difference instabilities,
Phys. Rev. Accel. Beams 20, 104201 (2017).
\bibitem{ingo4}I. Hofmann and O. Boine-Frankenheim, Revisiting the
Longitudinal 90° Limit in High Intensity Linear Accelerators, Phys. Rev. Lett. 118, 
	114803, (2017).
\bibitem{ingoprab17}I. Hofmann and O. Boine-Frankenheim, 
Parametric instabilities in 3D periodically focused beams with space
charge, Phys. Rev. Accel. Beams 20, 014202 (2017).
\bibitem{qiangenv}J. Qiang, Three-dimensional envelope instability in periodic focusing channels, Phys. Rev. Accel. Beams 21,
034201 (2018).
\bibitem{qiangenv2}J. Qiang, Mitigation of envelope instability through fast acceleration
in linear accelerators, Phys. Rev. Accel. Beams 21,
114201 (2018).
\bibitem{ryne}{R. Ryne, Finding matched rms envelopes in rf linacs: A Hamiltonian approach, \url{https://arxiv.org/abs/acc-phys/9502001}}
\bibitem{sacherer}F. J. Sacherer, RMS envelope equations with space charge,
IEEE Trans. Nucl. Sci. 18, 1101 (1971).
\bibitem{qiangipac25}J. Qiang, Y. Hao, A. Qiang, J. Wan, A module for fast auto differentiable simulations, in Proc. of IPAC25, WEBN2, p.1671, 2025. \url{https://github.com/qianglbl/mAD}
\bibitem{adwiki}\url{https://en.wikipedia.org/wiki/Automatic\_differentiation}

\end{thebibliography}
\end{document}